\documentclass{pasj00}
\begin{document}
\SetRunningHead{S.\ Yamauchi et al.}{X-Ray Emission from the Galactic Supernova Remnant G12.0$-$0.1}
\Received{yyyy/mm/dd}%{yyyy/mm/dd}
\Accepted{yyyy/mm/dd}%{yyyy/mm/dd}

\title{X-Ray Emission from the Galactic Supernova Remnant G12.0$-$0.1}

%%% begin:list of authors
\author{Shigeo \textsc{Yamauchi}}%
\affil{Department of Physics, Faculty of Science, Nara Women's University, Kitauoyanishi-machi, Nara 630-8506}
\email{yamauchi@cc.nara-wu.ac.jp}

\author{Aya \textsc{Bamba}}
\affil{Department of Physics and Mathematics, Aoyama Gakuin University, \\
5-10-1 Fuchinobe, Chuo-ku, Sagamihara, Kanagawa 252-5258}
\and 
\author{Katsuji \textsc{Koyama}}
\affil{Department of Physics, Graduate School of Science, Kyoto University, 
Kitashirakawa-oiwake-cho, Sakyo-ku, Kyoto 606-8502\\
Department of Earth and Space Science, Graduate School of Science, Osaka University, \\
1-1 Machikaneyama-cho, Toyonaka, Osaka 560-0043}

%% `\KeyWords{}' always has to be placed before `\maketitle'.
 %Do NOT move this preamble from here!

\maketitle

\begin{abstract}
We present results of the Suzaku/XIS observation around the radio supernova remnant (SNR) 
G12.0$-$0.1.
No significant diffuse emission extending in or along the radio shell was observed.
Instead two compact X-ray sources, Suzaku J181205$-$1835 and Suzaku J181210$-$1842, 
were found in or near G12.0$-$0.1.
Suzaku J181205$-$1835 is located at the northwest of the radio shell of G12.0$-$0.1.
The X-ray profile is slightly extended over the point spread function of the Suzaku telescope.
The X-ray spectrum has no line-like structure and is well represented by a power-law model with 
a photon index of 2.2 and an absorption column of $N_{\rm H}$=4.9$\times$10$^{22}$ cm$^{-2}$.
The distances of Suzaku J181205$-$1835 and G12.0$-$0.1 are estimated from the absorption 
column and the $\Sigma$-$D$ relation, respectively, 
and are nearly the same with each other.
These results suggest that Suzaku J181205$-$1835 is a candidate of a pulsar wind nebula 
associated with G12.0$-$0.1.
From its location, Suzaku J181210$-$1842 would be unrelated object to G12.0$-$0.1.
The X-ray profile is point-like and the spectrum is thin thermal emission with 
Fe K-lines at 6.4, 6.7, and 6.97 keV, similar to those of cataclysmic variables. 
\end{abstract}

\section{Introduction}

Supernovae remnants (SNRs) are main sites of heavy element production and 
of acceleration of high-energy particles in the Galaxy.
X-rays are powerful tools to investigate the elements (abundances) 
by the plasma diagnostics and search for synchrotron X-rays from high-energy electrons.
Although 273 Galactic SNRs have been discovered so far \citep{Green2009} by the radio,
only limited fraction have been detected in the X-ray; X-ray properties of 
most of the SNRs have not been investigated yet.
We have performed the Galactic plane survey project with ASCA 
(ASCA Galactic plane survey, AGPS, \cite{Yamauchi2002}).
AGPS covered  the Galactic inner disk ($| l | < 45^{\circ}$ and $| b | < 0.4^{\circ}$) and the 
Galactic center region ($| l |<2^{\circ}$ and $| b |<2^{\circ}$)
with successive pointing  observations of about 10 ks exposure.
AGPS detected X-ray emissions from $\sim$30 radio cataloged SNRs, in which
15 SNRs were new discoveries in the X-ray  \citep{Sugizaki2001, Sakano2002, Yamauchi2002}.  
Among them, AGPS found an weak and diffuse X-ray source near G12.0$-$0.1,  
designated  as AX J181211$-$1835 \citep{Sugizaki2001}.
The X-ray spectrum of AX J181211$-$1835 was well represented by either a thin thermal 
emission or a power-law model \citep{Yamauchi2008}.
However, due to the poor photon statistics, the estimated parameters had large errors, 
and hence the origin of the X-ray was not well constrained. 

Since Suzaku has a better spectral resolution, wider energy band, 
and lower/more stable intrinsic background than the previous X-ray satellites \citep{Mitsuda2007}, 
it is the suitable instrument to study a faint and diffuse sources.
Based on the Suzaku data, \citet{Sezer2010} 
reported that the spectra of G12.0$-$0.1 were represented by a thermal emission$+$power-law model.
Since G12.0$-$0.1 is a faint source at the Galactic Ridge,
where a strong X-ray emission, called Galactic Ridge X-ray Emission (GRXE), is prevailing, 
the careful background subtraction is required.
Thus, we reanalyzed the Suzaku data paying particular concern 
to background subtraction.
In this paper, we report new results of the 
X-ray spectra and discuss the nature of G12.0$-$0.1.

\section{Observations and Data Reduction}

%%
% Figures 1
%%
\begin{figure*}
  \begin{center}
    \FigureFile(16cm,8cm){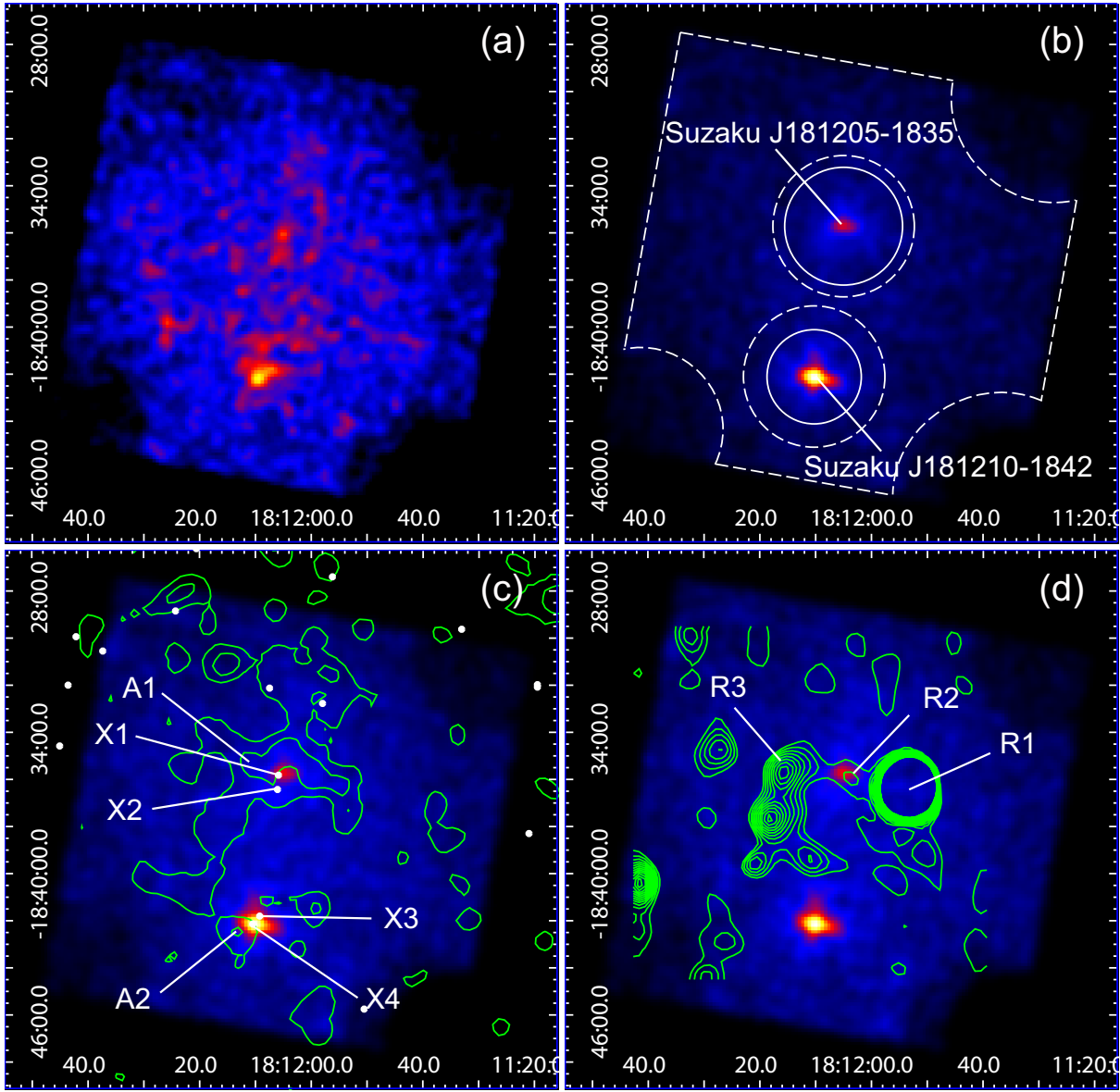}
  \end{center}
\caption{XIS images in the (a) 0.7--2, (b) 2--8, and  (c), (d) 0.7--8 keV 
smoothed with a Gaussian distribution of $\sigma$=24$''$.
The coordinates are J2000.0.  The data of XIS 0, 1, and 3 were co-added.
Neither background subtraction nor vignetting correction is performed.
The intensity levels of the X-ray and the radio bands are linearly spaced.
The contour in (c) shows the ASCA GIS2$+$3 intensity map in the 0.7--10 keV
smoothed with a Gaussian distribution of $\sigma$=60$''$, while the white dots
show the positions of X-ray sources in the XMM-Serendipitous Source Catalog
\citep{Watson2009}.
A1 and A2 are the ASCA source named as AX J181211$-$1835 and AX J181213$-$1842, respectively, 
while X1, X2, X3, and X4 are XMM-Newton/Chandra sources named as 2XMM J181205.8$-$183549, 
2XMM J181206.0$-$183625, 
2XMM J181209.2$-$184149=CXO J181209.1$-$184146, and 
2XMM J181210.5$-$184208=CXO J181210.3$-$184208, respectively.
The contour in (d) shows the radio intensity map at 1.4 GHz using the NRAO 
VLA Sky Survey (NVSS) \citep{Condon1998}. 
R1, R2, R3 are G11.944$-$0.037, NVSS J181203$-$183558, and G12.0$-$0.1, respectively.
The solid and dashed lines in (b) show source and background regions in the Suzaku analysis, 
respectively.
 }\label{fig:sample}
\end{figure*}
%%%

Suzaku observed the G12.0$-$0.1 region on 2007 October 2--3 with the X-ray CCD cameras (XIS)
on the focal planes of the thin foil X-ray Telescopes (XRT).  
Details of Suzaku, XIS, and XRT are given in  \citet{Mitsuda2007}, \citet{Koyama2007},
and \citet{Serlemitsos2007}, respectively.
XIS sensor-1 (XIS 1) is a back-side illuminated CCD (BI), while
the other three XIS sensors (XIS 0, 2, and 3) 
are front-side illuminated CCDs (FIs).
The observed field was 17.$'$8$\times$17.$'$8 area with the center at 
($\alpha$, $\delta$)$_{\rm J2000.0}$= (\timeform{273.D0208}, \timeform{-18.D6250}).
Since one of the FIs  (XIS2) turned dysfunctional in 2006 November, 
we used the data obtained with XIS 0, XIS 1, and XIS 3.
The XIS was operated in the normal clocking mode with the time resolution of 8 s.
The XIS employs the spaced-row charge injection (SCI) technique to rejuvenate 
the spectral resolution. Details concerning on the SCI technique are given in \citet{Nakajima2008} 
and \citet{Uchiyama2009}.

Data reduction and analysis were made using the HEAsoft 
version 6.11.
The XIS pulse-height data for each X-ray event were converted to 
Pulse Invariant (PI) channels using the {\tt xispi} software version 
2009-02-28 and the calibration database version 2011-11-09.
We excluded the data obtained at the South
Atlantic Anomaly, during the earth occultation, and at the low elevation
angle from the earth rim of $<$ 5$^{\circ}$ (night earth) and $<20^{\circ}$
(day earth).
After removing hot and flickering pixels,
we used the grade 0, 2, 3, 4, and 6 data.
The resultant exposure time was 53.7 ks for each XIS detector.

\section{Analysis and Results}

\subsection{Image}

%%
% Figures 2
%%
\begin{figure}
  \begin{center}
    \FigureFile(8cm,8cm){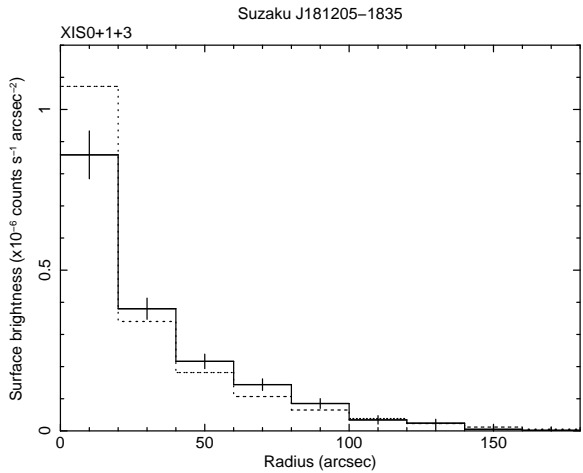}
    %%% \FigureFile(width,height){filename}
  \end{center}
  \caption{Radial profile of Suzaku J181205$-$1835 (the solid line) in the 2--8 keV energy band 
  and the best-fit PSF (the dotted line).
  The background counts estimated from an annulus region of 180$''$--210$''$ radii
were subtracted.
}\label{fig:sample}
\end{figure}

Figure 1 shows X-ray images in the 0.7--2, 2--8, and 0.7--8 keV energy bands.
The data of XIS 0, 1, and 3 were added to increase photon statistics.
In the energy band above 2 keV (figure 1b), we see two compact X-ray sources.  
The north and the south sources have peaks at 
(RA, Dec)$_{\rm J2000.0}$=(18$^{\rm h}$12$^{\rm m}$05.0$^{\rm s}$,
$-$18$^{\circ}$35$'$43$''$) and 
(RA, Dec)$_{\rm J2000.0}$=(18$^{\rm h}$12$^{\rm m}$10.3$^{\rm s}$,
$-$18$^{\circ}$42$'$07$''$), and hence we named them Suzaku J181205$-$1835 
and Suzaku J181210$-$1842, respectively.
The typical positional uncertainty of Suzaku is 19$''$ \citep{Uchiyama2008}.
In addition, the systematic error of the peak determination is 8$''$.

\subsection{Suzaku J181205$-$1835}

In the 0.7--8keV band image of figure 1c, we show the ASCA intensity profile (the contour)  
and the peak position (labels A1 and A2), while the XMM-Newton/Chandra sources 
are displayed as X1, 2, 3, and 4. 
The error circle of Suzaku J181205$-$1835 includes 2XMM J181205.8$-$183549 (label X1) 
(the Second XMM-Newton Serendipitous Source Catalog Third Data Release: 
2XMMi-DR3,  \cite{Watson2009}), and hence
Suzaku J181205$-$1835 is identified with 2XMM J181205.8$-$183549.
The nearest ASCA source is AX J181211$-$1835 (label A1) (the separation angle: 
$\Delta \theta$=1$'$.58).  
Since AX J181211$-$1835 is a faint source (5.4$\sigma$ detection 
in the 0.7--7 keV band; \cite{Sugizaki2001}) and  is elongated to the east-west 
direction (see figure 1c), the uncertainty of the peak position would be large.  
Thus, Suzaku J181205$-$1835 and AX J181211$-$1835 are probably identical.

The radio intensity map at 1.4 GHz \citep{Condon1998} is overlaid in the Suzaku image of figure 1d. 
An weak radio source NVSS J181203$-$183558 (label R2, NED; \cite{Condon1998}) is near 
to Suzaku J181205$-$1835 ($\Delta \theta$=23$''$) and their error regions overlap each other.  
However, the separation angle between 2XMM J181205.8$-$183549 and NVSS J181203$-$183558 
($\Delta \theta$=32$''$) is larger than the positional errors of XMM-Newton (5$''$.22 and systematic 
error=1$''$, 2XXMi-DR3) and NVSS (19$''$, NED). 
Thus, Suzaku J181205$-$1835 may not be identified with NVSS J181203$-$183558.
We also searched for optical and infrared counterparts for Suzaku J181205$-$1835 by using the SIMBAD 
database and found no candidate.

Figure 2 shows the radial profiles in the 2--8 keV band, Suzaku J181205$-$1835 (the solid line) 
and the best-fit Point Spread Function (PSF, the dotted line) at the source position made using {\tt xissim}.  
The background counts, estimated from an annulus region with radii of 180$''$--210$''$, 
were subtracted.
A free parameter was a normalization of the PSF.
The observed profile cannot be reproduced by the PSF with 
$\chi^2$/ d.o.f.=18.49/8=2.31; it is slightly extended over the PSF.
We note that 2XMM J181205.8$-$183549, an XMM-Newton counterpart, is also slightly 
extended (15$''$.7, 2XMMi-DR3).

%%
% Figures 3
%%
\begin{figure}[t]
  \begin{center}
    \FigureFile(8cm,8cm){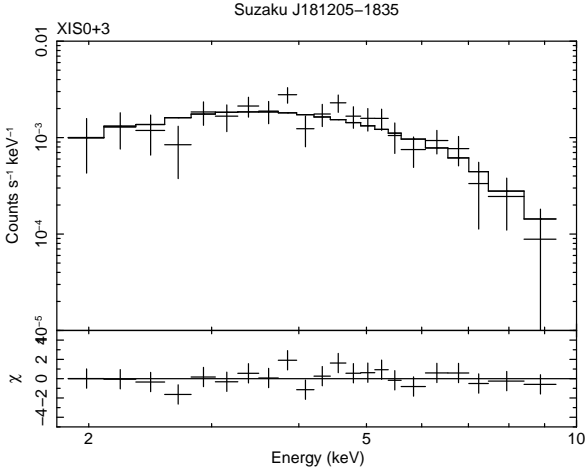}
    %%% \FigureFile(width,height){filename}
  \end{center}
  \caption{X-ray spectrum of Suzaku J181205$-$1835 (XIS 0$+$3).
The histogram shows the best-fit power-law model (see table 1).
}\label{fig:sample}
\end{figure}

%%%%%%
% Table 1
%
\begin{table}[t]
\caption{The best-fit parameters for the spectrum of Suzaku J181205$-$1835.$^{\ast}$ }
\begin{center}
\begin{tabular}{ll} \hline \\ [-6pt]
Parameter &  Value \\
\hline \\[-6pt]
\multicolumn{2}{c}{Model: power-law$\times$absorption}\\
\hline \\[-6pt]
$N_{\rm H}$ ($\times10^{22}$ cm$^{-2}$) & 4.9$^{+3.0}_{-2.3}$\\
$\Gamma$ & 2.2$^{+0.8}_{-0.7}$\\
$\chi^2$/d.o.f. & 31.36/39\\
Flux$^{\dag}$ ($\times10^{-13}$ erg s$^{-1}$ cm$^{-2}$) & 3.0$\pm$0.4\\
\hline \\[-6pt]
\end{tabular}
\end{center}
\vspace{-0.3cm}
$^{\ast}$ Errors are single parameter 90\% confidence levels ($\Delta \chi^2$ $<$ 2.7).\\
$^{\dag}$ Observed (absorbed) flux in the 2--10 keV energy band calculated from the best-fit model.\\
\end{table}
%%%%%%

X-ray spectra of Suzaku J181205$-$1835 
were extracted from a 2$'$.5 radius circle (see figures 1 and 2).
The background spectra were extracted from a source free region in the same FOV, 
as shown in figure 1b.
We constructed the non-X-ray background (NXB) for the source and the background spectra
from the night earth data using {\tt xisnxbgen} \citep{Tawa2008}.
After subtraction of the NXB from the source and the background spectra, 
we corrected the vignetting effect of the background spectra
by the method shown by \citet{Hyodo2008} and subtracted 
the vignetting-corrected background spectra from the source spectra.
The X-ray counts in the 1--10 keV energy band extracted from the source region 
were 1967, 2625,  and 1992 for XIS 0, 1, and 3, respectively, while 
after the background subtraction, the source counts were 467, 366, and 448, respectively. 
We co-added the XIS 0 and XIS 3 spectra, but treated the XIS 1 spectrum 
separately, because the response functions of the FIs and BI are different.
Then, the background-subtracted spectra were grouped with 
a similar signal-to-noise ratio for a bin.
Response files, Redistribution Matrix Files (RMFs) and Ancillary Response Files (ARFs), 
were made using {\tt xisrmfgen} and {\tt xissimarfgen}, respectively.

The background-subtracted spectrum is shown in figure 3.
Only the FI CCD spectrum (XIS 0$+$3) is displayed for brevity.
The X-ray spectra exhibit no emission line feature. 
We thus simultaneously fitted the XIS 0$+$3 and the XIS 1 spectra with 
a power-law model, modified by low-energy absorption.
The cross sections of photoelectric absorption were taken from 
Balucinska-Church and McCammon (1992).
This model can explain the spectra well with  $\chi^2$/ d.o.f.=31.16/39=0.80.
The best-fit parameters are listed in table 1, 
while the best-fit power-law model is shown in figure 3.

Since  ASCA reported  more extended emission \citep{Yamauchi2008}, 
we made Suzaku spectra within a  4$'$.5  circle from the source 
(the same size for the ASCA spectrum). 
The background spectra were extracted from the outer source free region in the same FOV.  
Then, the flux in the 2--10 keV band is 
(3.5$\pm$0.6)$\times$10$^{-13}$ erg s$^{-1}$ cm$^{-2}$, nearly the same as that from
the  2$'$.5 circle  region (see table 1).
Thus, we conclude that there is no largely extended non-thermal X-ray emission 
as reported with ASCA\footnote {We made the same analysis with no vigneting correction 
for the background spectra, 
and found that the flux is (8.0$\pm$0.8)$\times$10$^{-13}$ erg s$^{-1}$ cm$^{-2}$,
larger than that in table 1.
Thus, we suspect that the ASCA extended emission is artifact caused mainly 
by an improper background subtraction (no veigenting correction).}.

We also examined possibility of largely extended thermal X-ray emission, 
by adding a thermal model with solar abundances \citep{Anders1989} ({\tt apec} model in XSPEC) 
to the power-law function 
fixing the best-fit photon index and $N_{\rm H}$ value to those in table 1.
We scanned the temperature range of 0.2--1 keV, and then obtained 
an upper limit of the thermal emission 
with $kT<$1 keV
(absorbed) within a 4$'$.5 radius circle to be 6$\times$10$^{-14}$ erg s $^{-1}$ cm$^{-2}$  
in the 2--10 keV band (90\% confidence level). 

We made timing data (X-ray counts/8 s) from a circle with a radius of 1$'$.5 
centered on the peak position of Suzaku J181205$-$1835 for each detector and 
co-added the XIS 0, 1, and 3 data to maximize photon statistics.
We found no significant intensity variation in the X-ray light curve.
We also searched for a pulsation using {\tt powspec}, but found 
no clear coherent pulsation in the range of 16--2048 s.

\subsection{Suzaku J181210$-$1842}

As shown in figure 1c,
the position of the south source, Suzaku J181210$-$1842, well coincides with that of
2XMM J181210.5$-$184208=CXO J181210.3$-$184208
(label X4, \cite{Cackett2006,Evans2010}; 2XMMi-DR3). 
The error circle of Suzaku J181210$-$1842 also overlaps with that of 
AX J181213$-$1842 ($\Delta \theta$=0$'$.90).
Therefore, Suzaku J181210$-$1842 is identified with 
2XMM J181210.5$-$184208=CXO J181210.3$-$184208=AX J181213$-$1842.
No radio counterpart was found at the position of Suzaku J181210$-$1842 (figure 1d).
On the other hand, 3 optical and infrared sources within the error region, 
USNO-B1.0 0712$-$00536743, 2MASS J18121061$-$184233, 
and SSTGLMC G0118947$-$00.1396, were found in the SIMBAD database.

Using the same method as that for Suzaku J181205$-$1835,
we made a radial profile of Suzaku J181210$-$1842 and compared  with the PSF.
Figure 4 shows the radial profile of Suzaku J181210$-$1842 and the best-fit PSF.
The observed profile is well fitted with the PSF  ($\chi^2$/d.o.f.=11.41/8=1.43), 
and hence Suzaku J181210$-$1842 is a point source.
We note that 2XMM J181210.5$-$184208, an XMM-Newton counterpart, is also
a point source (2XMMi-DR3).

%%
% Figures 4
%%
\begin{figure}
  \begin{center}
    \FigureFile(8cm,8cm){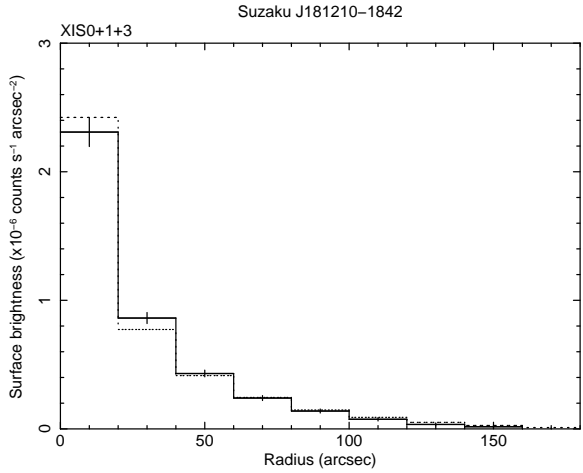}
    %%% \FigureFile(width,height){filename}
  \end{center}
  \caption{Same as figure 2, but for Suzaku J181210$-$1842.
}\label{fig:sample}
\end{figure}

X-ray spectra of Suzaku J181210$-$1842 
were extracted from a 2$'$.0 radius circle with the same background subtraction process 
as that for Suzaku J181205$-$1835.
The X-ray counts from the source region in the 1--10 keV energy band 
were 1921, 2081, and 1660  for XIS 0, 1, and 3, respectively, while
after the background subtraction the source counts were 969, 694, and 835, 
respectively.
Then, the background-subtracted spectra were grouped 
as same as Suzaku J181205$-$1835.

Figure 5 shows the background-subtracted spectrum.
Only the XIS 0$+$3 spectrum is displayed for brevity.
We found a hump near the energy of Fe K-lines.
We simultaneously fitted the XIS 0$+$3 and the XIS 1 spectra 
with a model consisting of thermal bremsstrahlung and a Gaussian line, 
and found a very broad line ($\sigma \sim230$ eV) leaving a wavy residual 
(the $\chi^2$ value of 85.23 for d.o.f.=75).
This suggests that the hump consists of several emission lines.
Therefore, we next fitted the spectra adding three narrow Gaussian lines at 6.4, 6.7, 
and 6.97 keV to the thermal bremsstrahlung model. The center energies of the two lines 
were fixed at 6.4 and 6.97 keV. 
The fit was improved (the $\chi^2$ value of 80.61 for d.o.f.=74).
The best-fit parameters are listed in table 2 and the best-fit model is displayed in figure 5.

We extracted X-ray counts in the 1--8 keV band from a 1$'$.5 radius circle, co-added the XIS 0, 1, 
and 3 data, and examined time variation.
No significant intensity variation was found in the X-ray light curve.
We also searched for a pulsation but found no clear pulsation in the range of 16--2048 s.

%%
% Figures 5
%%[
\begin{figure*}[t]
  \begin{center}
    \FigureFile(8cm,8cm){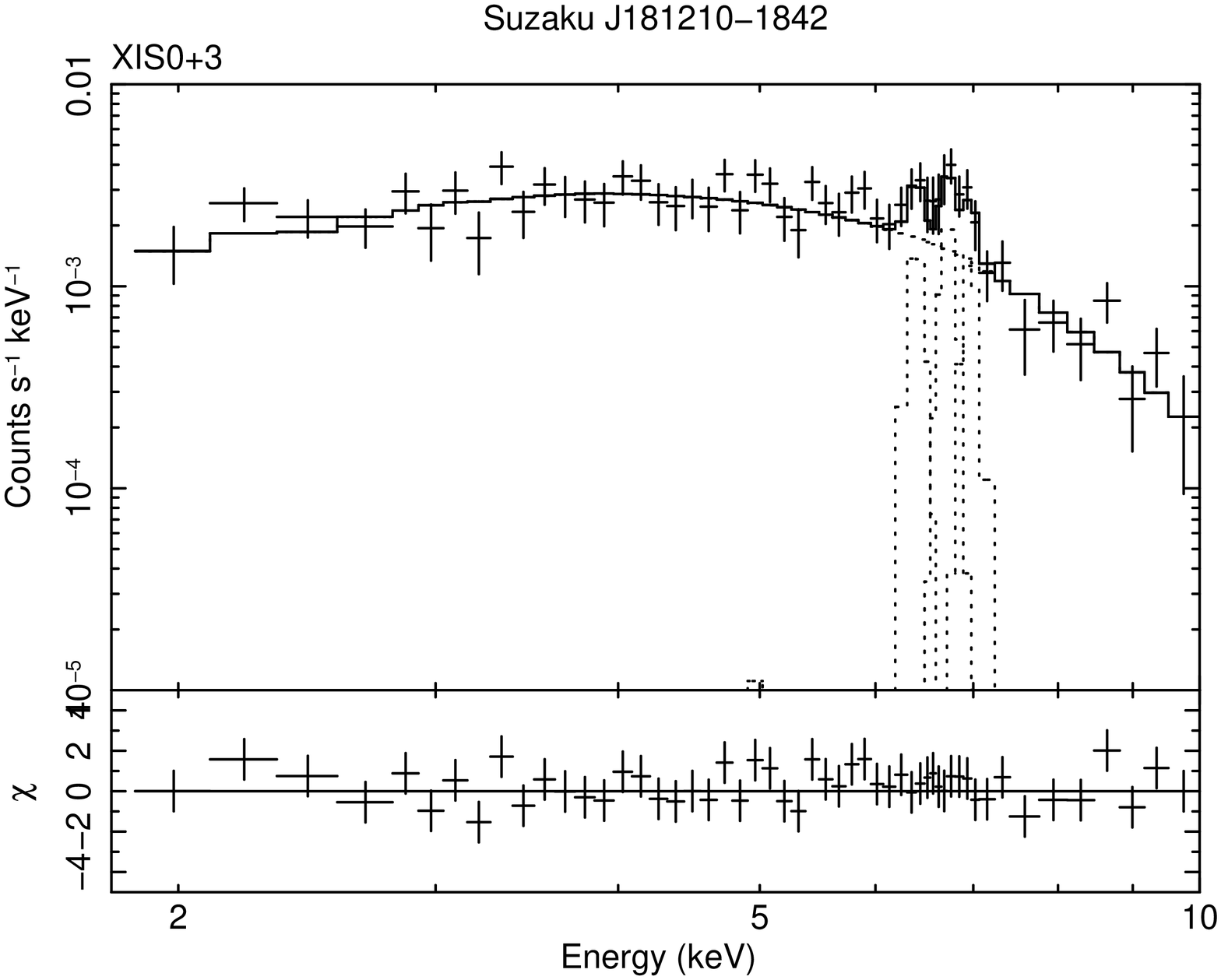}
   \FigureFile(8cm,8cm){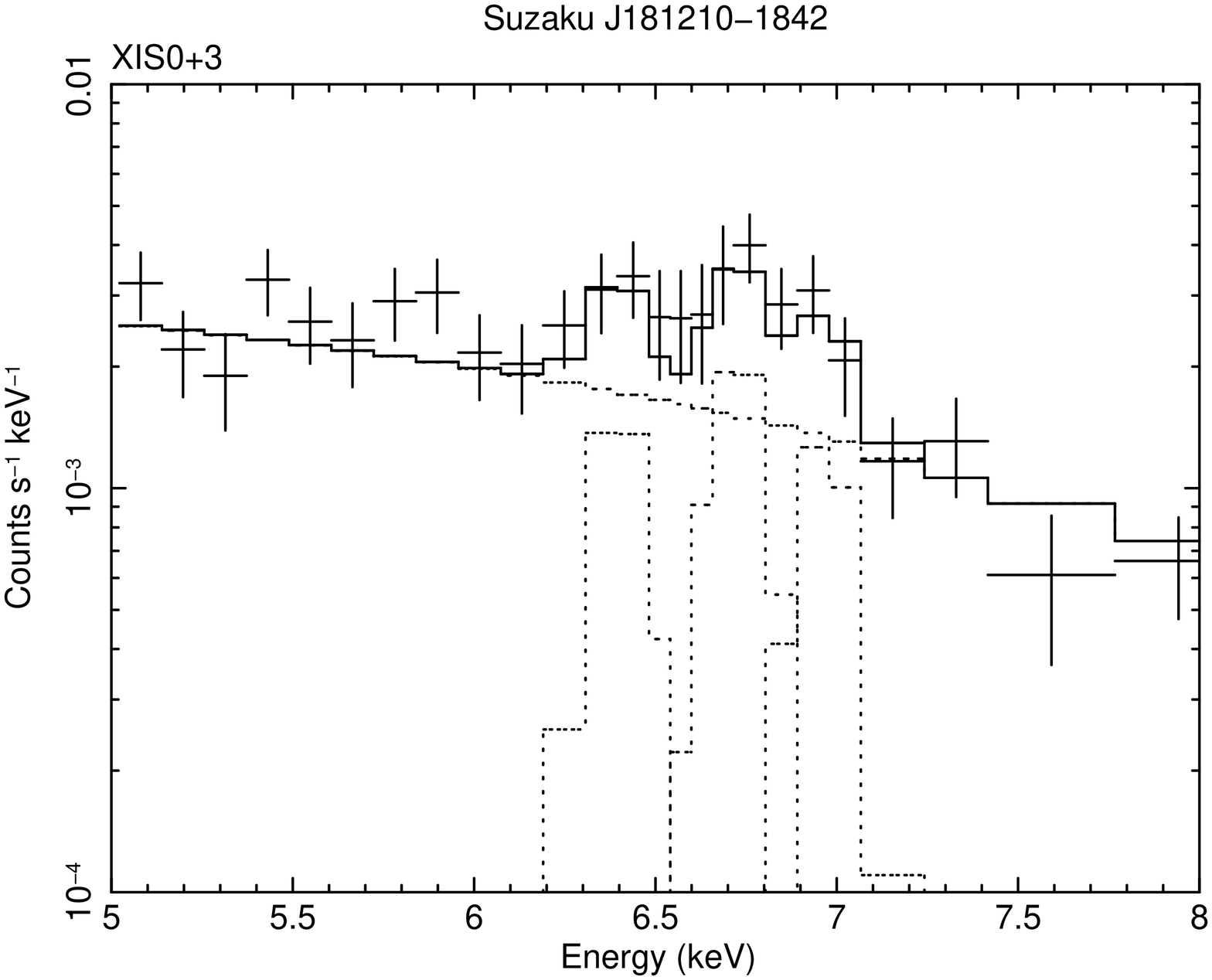}
    %%% \FigureFile(width,height){filename}
  \end{center}
  \caption{Left: 
XIS spectrum (XIS 0$+$3) of Suzaku J181210$-$1842  (upper panel) 
and residuals from the best-fit model (lower panel).
The histogram shows the best-fit bremsstrahalung $+$ emission lines model (see table 2).
Right: 
XIS spectrum (XIS 0$+$3) near the Fe-K energy (5--8 keV band).
The histogram shows the best-fit model.
}\label{fig:sample}
\end{figure*}

%%%%%%
% Table 2
%
\begin{table}[t]
\caption{The best-fit parameters for the spectrum of Suzaku J181210$-$1842.$^{\ast}$ }
\begin{center}
\begin{tabular}{ll} \hline \\ [-6pt]
Parameter &  Value \\
\hline \\[-6pt]
\multicolumn{2}{c}{Model: (bremsstrahalung$+$emission lines)$\times$absorption}\\
\hline \\[-6pt]
$N_{\rm H}$ ($\times10^{22}$ cm$^{-2}$) & 3.4$^{+0.8}_{-0.6}$ \\
$kT$ (keV) & $>$38\\
$E_{\rm Line1}^{\dag}$ (keV)  & 6.4 (fixed)\\
$EW_{\rm Line1}^{\ddag}$ (eV) & 175$\pm$90\\
$E_{\rm Line2}^{\dag}$ (keV)  & 6.73$^{+0.06}_{-0.05}$ \\
$EW_{\rm Line2}^{\ddag}$ (eV) & 270$\pm$110\\
$E_{\rm Line3}^{\dag}$ (keV)  & 6.97 (fixed)\\
$EW_{\rm Line3}^{\ddag}$ (eV) & 195$\pm$110\\
$\chi^2$/d.o.f. & 80.61/74\\
Flux$^{\S}$ ($\times10^{-13}$ erg s$^{-1}$ cm$^{-2}$) & 8.7$\pm$0.5\\
\hline \\[-6pt]
\end{tabular}
\end{center}
\vspace{-0.3cm}
$^{\ast}$ Errors are single parameter 90\% confidence levels ($\Delta \chi^2$ $<$ 2.7).\\
$^{\dag}$ Energy of the emission line.\\
$^{\ddag}$ Equivalent width of the emission line.\\
$^{\S}$ Observed (absorbed) flux in the 2--10 keV energy band calculated from the best-fit model.\\
\end{table}
%%%%%%

\section{Discussion}

Contrary to the ASCA result, Suzaku found neither thin thermal nor non-thermal X-rays 
extending largely in or along the radio shell of the SNR G12.0$-$0.1.
Instead two compact X-ray sources, Suzaku J181205$-$1835 and Suzaku J181210$-$1842, 
were found in and near G12.0$-$0.1.  
Although these two sources have been already reported with XMM-Newton, 
the Suzaku observation provides more accurate spectra and more severe constraint.    
Here, we discuss the nature of these sources based on the Suzaku results.

\subsection{Suzaku J181205$-$1835}

The Galactic HI column density ($N_{\rm HI}$) along the line-of-sight to
Suzaku J181205$-$1835 is $N_{\rm HI}$=2.0$\times$10$^{22}$ cm$^{-2}$ \citep{Dickey1990} 
or $N_{\rm HI}$=1.4$\times$10$^{22}$ cm$^{-2}$ \citep{LAB}.  
Using the CO intensity at the source position (164 K km s$^{-1}$; \cite{Dame2001}) 
and the conversion factor to the $N_{\rm H_2}$ value 
(1.8$\times$10$^{20}$ cm$^{-2}$ K$^{-1}$ km$^{-1}$ s; \cite{Dame2001}), 
the Galactic H$_2$ column density ($N_{\rm H_2}$) is estimated to be 
$N_{\rm H_2}$=3.0$\times$10$^{22}$ cm$^{-2}$.
The total $N_{\rm H}$, $N_{\rm H}$=$N_{\rm HI}$$+$2$N_{\rm H_2}$, is then 
(7.3--7.9)$\times$10$^{22}$ cm$^{-2}$.
The $N_{\rm H}$=4.9$\times$10$^{22}$ cm$^{-2}$ of Suzaku J181205$-$1835 is a bit larger than
the half of the total $N_{\rm H}$ thorough the Galaxy.
Therefore, Suzaku J181205$-$1835 would be in the Galactic inner disk at
the distance of $\sim$10 kpc.

No reliable estimation, nor any report for the distance to G12.0$-$0.1
has been available so far.  We therefore applied empirical relation
between the surface brightness and the SNR diameter 
($\Sigma$--D relation: e.g., \cite{Case1998,Pavlovic2013}). 
Then the distance of G12.0$-$0.1 is calculated to be $\sim$11 kpc .
Although the distance estimation methods for Suzaku J181205$-$1835 and
G12.0$-$0.1 are the subject of large uncertainty, 
the distances of the two objects are roughly the same.
We hence assume that Suzaku J181205$-$1835 is associated with 
SNR G12.0$-$0.1.
In the following discussion, we assume the distance to 
Suzaku J181205$-$1835/G12.0$-$0.1 to be 10 kpc, with the reservation
of a large error.

Suzaku J181205$-$1835 is spatially extended by sub-arcminute. 
The X-ray spectrum can be explained by a power-law model with a photon index of 
$\Gamma$=2.2$^{+0.8}_{-0.7}$, 
a typical value of SN 1006-like SNRs ($\Gamma\sim$2.5--3, e.g., \cite{Koyama1995, Bamba2005}) 
or pulsar wind nebulae (PWNe) ($\Gamma\sim$2, e.g., \cite{Possenti2002,Kargaltsev2008}).
No largely extended emission (power-law) near or around 
Suzaku J181205$-$1835 was found, which excludes possibility of SN 1006-like object.

The X-ray luminosity in the 2--10 keV band is calculated to be
5.6$\times$10$^{33}$ erg s$^{-1}$ at 10 kpc. 
Based on the empirical relation between an X-ray luminosity and 
a characteristic age of pulsars in PWNe \citep{Possenti2002,Mattana2009},
the age of Suzaku J181205$-$1835 is 
estimated to be $\sim$3$\times$10$^{3}$--3$\times$10$^4$ yr.
Middle-aged PWNe with characteristic ages of $\sim$3$\times$10$^{3}$--3$\times$10$^4$ yr
are expected to have small synchrotron nebulae with a size of $\sim$2--10 pc 
\citep{Bamba2010}. 
This corresponds to 0.$'$7--3$.'$4 at 10 kpc, in agreement with the extent of 
Suzaku J181205$-$1835 (figure 2).

In order to examine long-term intensity variation of Suzaku J181205$-$1835, 
we re-analyzed the ASCA spectrum from a 3$'$ radius circle 
after subtracting the background counts from an annulus region of 3$'$--6$'$ radii. 
The flux is then (3.7$\pm$2.3)$\times$10$^{-13}$ erg s$^{-1}$ cm$^{-2}$ (2--10 keV),  
the same as the Suzaku result (see table 1).
The X-ray flux of 2XMM J181205.8$-$183549 is estimated from the data 
of short exposure time (0.3--1.3 ks, 2XMMi-DR3). 
Although the values are different from detector to detector (MOS1, MOS2 and PN), 
they are in the range of (1.2--21)$\times$10$^{-13}$ erg s$^{-1}$ cm$^{-2}$.
The Suzaku result (3.4$\times$10$^{-13}$ erg s$^{-1}$ cm$^{-2}$ in the 2--12 keV band) 
is in this range.
Thus, we conclude that there is no significant long-term intensity variation 
among the ASCA (1996), XMM-Newton (2003), and Suzaku (2007) observations.

These characteristics, the distance, non-thermal X-rays, reasonable extent, and 
no significant long-term intensity variation, suggest that Suzaku J181205$-$1835 is 
a new PWN candidate associated with SNR G12.0$-$0.1.
The thermal X-ray flux near or around this PMN candidate is less than 20\% of the PWN flux,  
and hence G12.0$-$0.1 would be a PWN dominant SNR. 
If Suzaku J181205$-$1835 is a PWN, a fast rotation of a neutron star with a spin period 
less than 1 s would be expected.  
Since the XIS has no capability to detect such a fast rotation, 
we used the Hard X-ray Detector (HXD; \cite{Takahashi2007}) data onboard Suzaku 
for the fast timing analysis with a 61~$\mu$s time resolution and an accuracy of 
$1.9\times 10^{-9}$ s\ s$^{-1}$ per day \citep{Terada2008}.
The resulting timing analysis in the 10--25~keV HXD data, 
however, showed no significant pulsation in the scanned period range of 10~ms to 1~s.
Thus, future search for a pulsation in the radio and X-ray bands are encouraged.

Suzaku J181205$-$1835 has a large offset from the center of G12.0$-$0.1, which 
requires a very large projected velocity of $\sim$1000 km s$^{-1}$. 
This value is near to a higher end of transverse velocities of radio pulsars
(ATNF pulsar catalog\footnote{http://www.atnf.csiro.au/people/pulsar/psrcat/}), and hence
Suzaku J181205$-$1835 may have a bow-shock structure. 
Assuming that Suzaku J181205$-$1835 has the same wide band spectrum as Crab nebula,
we can expect the total flux density of $\sim$20 mJy at 1 GHz.
However, 
since Suzaku J181205$-$1835 has no clear radio counterpart (see section 3.2 and figure 1d) and
the flux density of nearby radio source, NVSS J181203$-$183558, is $\sim$5 mJy at 1.4 GHz 
(\cite{Condon1998}, NED), 
the flux density of Suzaku J181205$-$1835 would be less than 5 mJy.
To reveal these issues, observations with high spatial resolution and high sensitivity 
in the radio and X-ray bands are also required.

\subsection{Suzaku J181210$-$1842}

Judging from its location, 
Suzaku J181210$-$1842 would be unrelated object to G12.0$-$0.1.
The X-ray lines at 6.4, 6.7, and 6.97 keV with the equivalent widths (EWs) of 
175$\pm$90, 270$\pm$110, and 195$\pm$110 eV, respectively, are typical features of
cataclysmic variables (CVs) (e.g., \cite{Hellier1998,Ezuka1999}).
Thus, Suzaku J181210$-$1842 is likely to be a CV
although no clear coherent pulsation has not been found.
The observed energy flux 
is  similar to those observed with  ASCA and XMM-Newton, 
(3--10)$\times$10$^{-13}$ erg s$^{-1}$ cm$^{-2}$ (\cite{Sugizaki2001,Cackett2006}; 2XMMi-DR3).

This discovery of a new CV candidate on the Galactic plane
means that there are many hidden CVs in the Galaxy.
CV is proposed to be a prime candidate of the GRXE 
(e.g., \cite{Yuasa2012}), 
because the GRXE spectra also exhibit three Fe K-lines at 
energies of 6.4, 6.7, and 6.97 keV \citep{Ebisawa2008,Yamauchi2009}.
We find that the EW of the 6.4 keV line of Suzaku J181210$-$1842 is similar to GRXE ($\sim$150 eV),  
while that of the 6.7 keV line ($\sim$200--300 eV) is about a half of the GRXE 
(400--600 eV, e.g.,  \cite{Yamauchi2009,Uchiyama2013}).
This statement is true for most of the CVs, which are bright
enough to measure the Fe K-lines (e.g., \cite{Ezuka1999}).
Thus, another population with a strong 6.7 keV line (EW $>$ 400 eV at least) is  required.
Although active binary stars (ABs) are also proposed to be a potential candidate 
of the GRXE \citep{Revnivtsev2009}, 
the 6.7 keV line is too weak to explain the GRXE Fe K-line features 
(e.g., \cite{Tsuru1989,Mewe1997,Guedel1999}).
Thus, if the discrete source origin is applied,
unresolved sources with a stronger Fe K-line than those of known bright CVs and ABs should exist.

\vspace{1pc}

The authors are grateful to all members of the Suzaku team. 
We thank Dr. Y. Terada for his useful comments.
This research made use of the NASA/IPAC Extragalactic Database (NED) 
operated by Jet Propulsion Laboratory, California Institute of Technology, 
under contract with NASA and 
the SIMBAD database operated at the CDS, Strasbourg, France. 
This work is supported in part by the Grant-in-Aid for Scientific
Research of the
Ministry of Education, Science, Sports and Culture 
(SY, No. 21540234 and 24540232).

%%%
% See the manual for the detail.
%%%

\end{document}